\documentclass[%
 reprint,
 amsmath,amssymb,
 aps,
]{revtex4-2}

\usepackage{graphicx}
\usepackage{dcolumn}
\usepackage{bm}
\usepackage{xcolor}%

\begin{document}

\preprint{APS/123-QED}

\title{Multiscale turbulence in stellarators}

\author{G. Merlo}
 \email{gabriele.merlo@ipp.mpg.de}

\author{A. {Bañón} Navarro}%
\author{T. G\"orler}%
\author{F. Jenko}%

\affiliation{%
  Max Planck Institute for Plasma Physics, Boltzmannstr. 2, Garching 85748, Germany
}

\author{F. Wilms}%
\affiliation{Proxima Fusion, Fl\"o{\ss}ergasse 2, 81369 Munich, Germany}

\date{\today}

\begin{abstract}
We present the first gyrokinetic simulations of multiscale turbulence in a stellarator, using the magnetic geometry of Wendelstein 7-X (W7-X) and experimentally relevant parameters. A broad range of scenarios is explored, including regimes where electron-temperature-gradient (ETG) turbulence coexists with varying levels of ion-temperature-gradient (ITG) turbulence, as well as cases involving microtearing modes (MTMs) relevant to high-$\beta$ and reactor-like conditions. Notably, while ETG turbulence does not form radial streamers as in tokamaks, it can still drive significant transport and interact with ion-scale turbulence. In electrostatic ITG-dominated regimes, electron-scale fluctuations erode zonal flows, enhancing ion-scale transport, while ion-scale turbulence suppresses ETG activity. In contrast, under electromagnetic MTM conditions, the isotropic nature of ETG turbulence limits its suppressive effect, allowing MTMs to persist. These findings underscore the critical role of cross-scale effects for accurate transport predictions in W7-X and future stellarators.
\end{abstract}

\maketitle

\textit{Introduction.} -- Magnetically confined fusion plasmas are a paradigmatic example of complex systems governed by multiscale and multiphysics processes. This is especially true for plasma microturbulence which usually dominates the radial transport of heat and particles and determines the energy confinement time, and  thus represents one of the main obstacles on the path towards achieving controlled fusion in a fusion power plant.

While plasma microturbulence research has traditionally focused on ion-scale dynamics, electron-scale turbulence driven by electron-temperature-gradient (ETG) modes can also play a significant role~\cite{Jenko_POP_2010,dorland_prl,jenko_2004, RenRevModPhys2024, Maeyama_NF_2024}. In tokamaks, ETG turbulence is known to drive experimentally relevant fluxes both in the core and in the pedestal region \cite{Kotschenreuther_2019,Howard_2016,Belli_2023,Maeyama2022,Hatch_2024}. However, the role of ETG turbulence in stellarators remains much less explored and understood.

More than two decades ago, pioneering studies \cite{Jenko_kendl_2002} have identified conditions under which ETG turbulence in the Wendelstein 7-AS stellarator can lead to the formation of streamers -- radially elongated vortices -- and experimentally relevant transport levels~\cite{dorland_prl}. On the other hand, it was argued in Ref.~\cite{Plunk_PRL_2019} that ETG transport may be negligible under reactor-like conditions. Meanwhile, follow-up studies have shown that ETG turbulence is relevant to present-day stellarator experiments. For example, Refs.~\cite{Weir_2021} and \cite{Wilms_NF_2024} have demonstrated that in W7-X Electron Cyclotron Resonance Heating (ECRH) scenarios, characterized by $T_e \sim T_i$ and $a/L_{T_e} > a/L_{T_i}$, including ETG transport was essential to reproduce the experimentally observed power balance. Further analysis across a broader set of conditions, including both Neutral Beam Injection (NBI) and mixed-heating scenarios~\cite{Fernando_PoP_2025}, confirmed that ETG-driven transport can play a significant role in certain parameter regimes.

In this Letter, we take a step toward bridging computational and experimental studies by presenting the first gyrokinetic simulations of multiscale turbulence in stellarator geometry, focusing on regimes dominated at ion scales by ion-temperature-gradient (ITG) modes and microtearing modes (MTMs). While cross-scale coupling is known to be important for such regimes in tokamaks, its relevance in stellarators has remained unclear. Here, we assess whether a scale-separated approach can reliably approximate the full multiscale dynamics in stellarator plasmas. Our results advance the understanding of turbulent transport in both present experiments and future fusion reactors.\\
\textit{Simulation Setup.} -- All simulations have been performed with the GPU-ported GENE code~\cite{Jenko_POP_2010,Germaschewski_PoP_2021}, which solves the nonlinear gyrokinetic equations. Plasma dynamics are described via the evolution of each species distribution function in the five-dimensional space $(x,y,z,v_\|,\mu)$, where $x,y,z$ are the radial, binormal, and parallel spatial coordinates; $v_\|$ the velocity parallel to the magnetic field; and $\mu$ is the magnetic moment. Although GENE is capable of modeling turbulence in a  stellarator retaining its entire 3D geometry~\cite{Maurer_JCP_2020}, the simulations discussed in this Letter are limited to the local limit, i.e., they model a narrow annulus around a given magnetic field line. This approximation, routinely applied to model core conditions, allows to use spectral methods in the perpendicular plane, whereas the generalized twist-and-shift boundary conditions~\cite{Martin_2018} are used in the parallel direction. Unless stated otherwise, the multiscale simulations use a computational box with $1536\times480\times80\times48\times12$ grid points for a spatial domain $L_x\times L_y=110\rho_s \times 65\rho_s$, where $\rho_s$ is the ion sound Larmor radius, thus resolving the range of binormal wave vectors $0<k_y\rho_s<46.3$.  The same radial box is also used for electron scale only simulations, with a larger value of $k_y^{min}$. The magnetic geometry is provided by the MHD solver GVEC~\cite{gvec_2025_zenodo} and reconstructed from W7-X experimental measurements. Collisions (modeled with a linearized Landau operator) and full electromagnetic perturbations (including $B_\|$ fluctuations) are retained.

Plasma parameters are taken from Ref.~\cite{Xanthopoulos_PRL_2020} describing W7-X plasma 20181016.037. We focus on the location $\rho_{tor}=0.4$ and on the line $\alpha=0$, where the ETG contribution was shown to be the largest~\cite{Wilms_NF_2024}. Thus, for the reference scenario we use $a/L_{T_e}=3.0$, $a/L_{T_i}=1.1$, $a/L_n=0.42$, and $T_i/T_e=0.75$. Here, $a$ is the average minor radius and $\rho_t=\sqrt{\psi/\psi_{\rm LCFS}}$, $\psi$ being the toroidal magnetic flux with its value at the last closed flux surface $\psi_{\rm LCFS}$ while $a/L\{T,n\}=-d\log(\{T,n\})/d\rho_t$ are normalized profile scale lengths. The experimental values of $\beta=0.38\%$ and collisionality $\nu=0.26\cdot10^{-3}$ are employed. As will be shown, the importance of cross-scale coupling depends on the relative strength of ion-to-electron-scale turbulence and zonal flows. Therefore, additional scenarios with different ITG drive will be considered as well, and parameter regimes in which the ITG modes are replaced by MTMs, as detailed below.

\begin{figure}[ht]
    \centering
    \includegraphics[width=\linewidth]{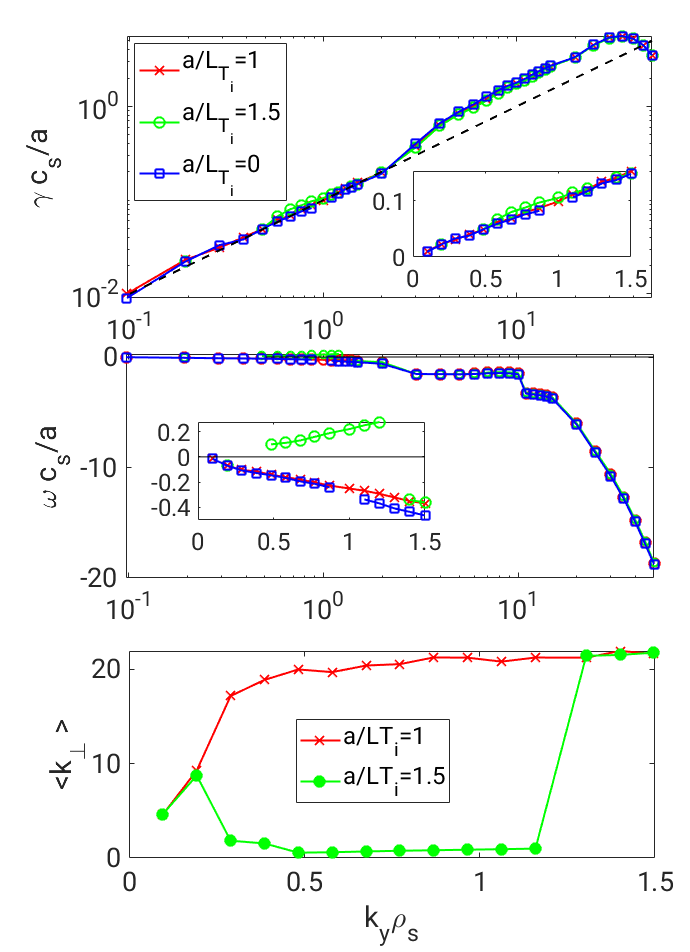}
    \caption{Linear growth rates (top), real frequencies (center), and eigenfunction averaged $\langle k_{\perp}\rangle$ (bottom) as functions of the binormal wavenumber, shown for different ion temperature gradient values. In the top panel, the dashed black line indicates the linear scaling $\gamma\propto k_y$.}
    \label{fig:1}
\end{figure}

\textit{Experimental Conditions.} -- We begin considering the experimental W7-X parameters, specifically $a/L_{T_i}=1.1$. The resulting linear growth rate and real frequency $k_y$-spectra are shown in Fig.~\ref{fig:1} as red crosses, together with the eigenfunction averaged $\langle k_{\perp}\rangle=\int\phi(z) k_{\perp}dz/\int\phi(z) dz$~\cite{Kotschenreuther_2019}. The linear results for other values of  $a/L_{T_i}=1.1$ are discussed below.
ETG modes (usually identified via negative real frequencies indicating electron diamagnetic drift direction and large values of  $\langle k_{\perp}\rangle$) dominate at all scales. At large scales, $k_y\rho_s<1$, ITG modes are unstable but are linearly subdominant as confirmed by the nonlinear results discussed below. Borrowing from tokamak literature~\cite{Staebler_Pop_2016}, we indicate with a dashed line the scaling  $\gamma\propto k_y$, which helps to gauge the importance of multiscale effects.
We present the first gyrokinetic simulations of multiscale turbulence in a stellarator, using the magnetic geometry of Wendelstein 7-X (W7-X) and experimentally relevant parameters. A broad range of scenarios is explored, including regimes where electron-temperature-gradient (ETG) turbulence coexists with varying levels of ion-temperature-gradient (ITG) turbulence, as well as cases involving microtearing modes (MTMs) relevant to high-$\beta$ and reactor-like conditions. Notably, while ETG turbulence does not form strong radial streamers as in tokamaks, it can still drive significant transport and interact with ion-scale turbulence. In electrostatic ITG-dominated regimes, electron-scale fluctuations erode zonal flows (similar to \cite{howard_multiscale_2016} in tokamaks), enhancing ion-scale transport, while ion-scale turbulence suppresses ETG activity. In contrast, under electromagnetic MTM conditions, the isotropic nature of ETG turbulence limits its suppressive effect, allowing MTMs to persist. These findings underscore the critical role of cross-scale effects for accurate transport predictions in W7-X and future stellarators.
\begin{figure}[h]
    \centering
    \includegraphics[width=\linewidth]{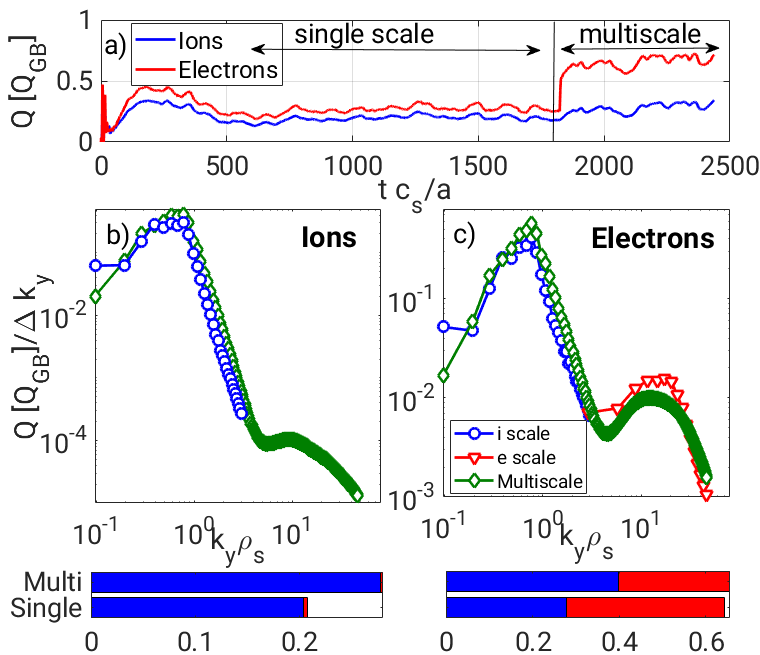}
    \caption{$(a)$ Time evolution of the $\mathbf{E\times B}$ heat fluxes for ions and electrons considering experimentally measured W7-X parameters. A single-scale low-$k_y$ simulation ($k_y\rho_s <3$) is run up to 1800 $a/c_s$, then continued with finer resolution ($k_y\rho_s<46$) to capture ETG scales. $(b,c)$ Comparison of heat flux spectra  from scale-separated and multiscale simulations for $(b)$ ions and $(c)$ electrons. Integrated flux values are shown below each plot: ion-scale contribution ($k_y\rho_s<3$) in blue, electron-scale contribution in red.}
    \label{fig:2}
\end{figure}
Figure \ref{fig:2}(a), shows the time traces of the nonlinear turbulent electron and ion energy fluxes in gyro-Bohm units, $Q_{\rm GB}=n_eT_ec_s \rho_s^2/a^2$ with ion sound speed $c_s$. The simulation is initiated as a single-scale low-$k_y$ run ($k_y^{\rm max}=3$) and then refined to include fine-scale ETGs. The initial state of the multiscale simulation is given by a snapshot of the single-scale one after saturation, with zero padding for the high-k modes. Therefore, the increase of the electron heat fluxes at $t=1800~a/c_s$ is attributable to the inclusion of ETG scales, whereas variations in the ion-scale flux as well as the differences in total electron transport compared to the sum of scale-separated results are consequences of cross-scale coupling. 

Focusing first on single-scale analyses, ion scales are characterized by a mixed ITG-TEM regime, driving similar fluxes through both channels ($Q_i/Q_{\rm GB}=0.2$, $Q_e/Q_{\rm GB}=0.27$) -- with zonal flows saturating the turbulence. ETGs at $k_y\rho_s >3$ drive an additional $Q_e/Q_{\rm GB}=0.36$, yielding a total electron flux of $Q_e/Q_{\rm GB}\sim 0.6$. The multiscale results provide a similar electron flux, $Q_e/Q_{\rm GB}=0.66$, but a $40\%$ larger ion transport. Inspecting the heat flux spectra in Figs.~\ref{fig:2}(b-c), we see how the increase of ion transport in the multiscale setup results from an increase at all scales. A similar increase at ion scales is also observed for $Q_e$. Therefore, the fact that the total electron heat flux is practically the same between single- and multi-scale runs is only coincidental, and the relative contribution from low and high $k_y$ varies between the two setups. Cross-scale coupling reduces the overall ETG contribution and increases the ion-scale one.

\begin{figure}
    \centering
    \includegraphics[width=\linewidth]{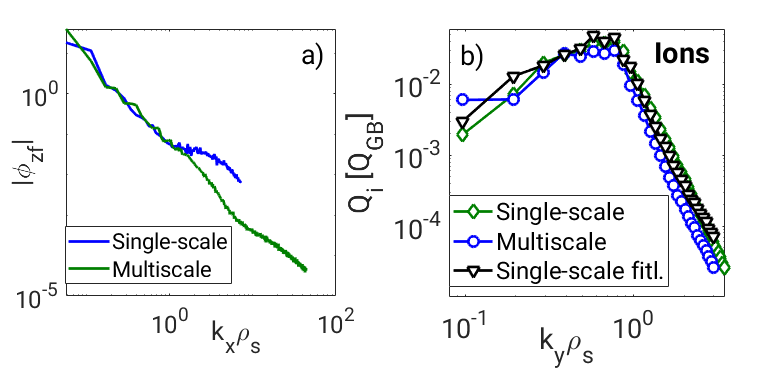}
    \caption{$(a)$ Radial spectra of zonal flow $\phi(k_y=0)$ showing erosion of high-k modes in multiscale versus single-scale simulations. $(b)$ Suppressing $k_x>2.5$ modes in single-scale recovers multiscale transport levels.}
    \label{fig:3}
\end{figure}

To understand the origin of the difference in the ion transport, in Figure \ref{fig:3}a, we show the radial spectra of the zonal flow $\phi_{\rm zf}=\int J(z)\phi(k_x,k_y=0,z)dz/\int J(z)dz$. One observes for $k_x\rho_s>2$ that the single-scale simulation shows a larger contribution  compared to the multiscale results. 
Furthermore, when artificially suppressing the zonal mode over the affected range (Fig.~\ref{fig:3}b) the same transport level as in a multiscale simulation is recovered, indicating that the increase of transport at ion scales is indeed a direct consequence of a different response to zonal flows.

\textit{Strong ITG Conditions.} -- In general, ion-scale turbulence driven by ITG modes may suppress electron-scale turbulence via shearing of small-scale eddies by larger-scale structures~\cite{Candy_2007,Goerler_PRL_2008,maeyama_cross-scale_2015,Howard_PoP_2016,Holland_2017,Howard_2018}. Conversely, electron-scale turbulence can either enhance or reduce ion-scale transport, depending on whether cross-scale coupling dampens the zonal flow or directly affects ion-scale instabilities~\cite{Maeyama_NF_2024}. To investigate whether similar behavior arises in stellarators, we consider a case with stronger ITG drive by setting $a/L_{T_i}=1.5$, a situation representative of strong NBI heating~\cite{Fernando_PoP_2025}.
\begin{figure}[h]
    \centering
    \includegraphics[width=\linewidth]{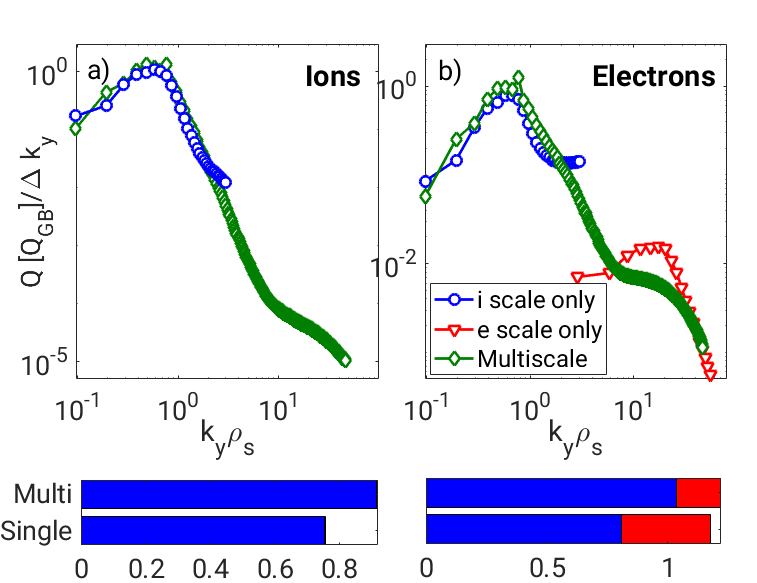}
    \caption{Ion $(a)$ and electron $(b)$ heat flux spectra from scale-separated and multiscale simulations with increased ITG drive $a/L_{T_i}=1.5$. Integrated fluxes are shown below each plot: ion-scale in blue, electron-scale in red.}
    \label{fig:4}
\end{figure}
Linear properties of this case are shown in Fig.~\ref{fig:1} (green circles). While the high-k part of the spectra remains unchanged, the range $0.5<k_y\rho_s< 1$ is now dominated by ITG modes as evidenced by the positive frequency of the most unstable mode. A significantly lower $\langle k_{\perp}\rangle$ is also observed, consistent with the ITG-dominated regime. Nonlinearly, as shown in Fig.~\ref{fig:4}, the turbulence exhibits a clear ITG character, with ion heat flux increasing by more than a factor of three and electron heat flux nearly doubling relative to the previous (weaker ITG) case.

Comparing single- and multiscale simulations, the latter shows a $20\%$ increase in the ion heat flux. This enhancement arises equally from cross-scale interactions (i.e., modification of ion-scale transport by small scales) and the inclusion of intermediate $2<k_y\rho_s<10$. Indeed the zonal field, not shown for brevity, is much less affected by the inclusion of high-k modes than in Fig.~\ref{fig:3}, indicating a diminished role of ETG modes under strong ITG drive. While the total electron heat flux remains similar between the sum of single-scale results and the multiscale simulation, this agreement is again coincidental. The ETG-scale ($k_y\rho_s>10$) contribution to $Q_e$ is reduced by a factor of four in the multiscale case and accounts for only $\sim10\%$ of the total. In contrast, the intermediate-k range contributes $20\%$ of $Q_e$.
These results suggest that the high-$k$ ETG turbulence becomes subdominant in reactor-relevant regimes with strong ITG drive, but only as a direct consequence of cross-scale coupling.

\begin{figure}[h]
    \centering
    \includegraphics[width=\linewidth]{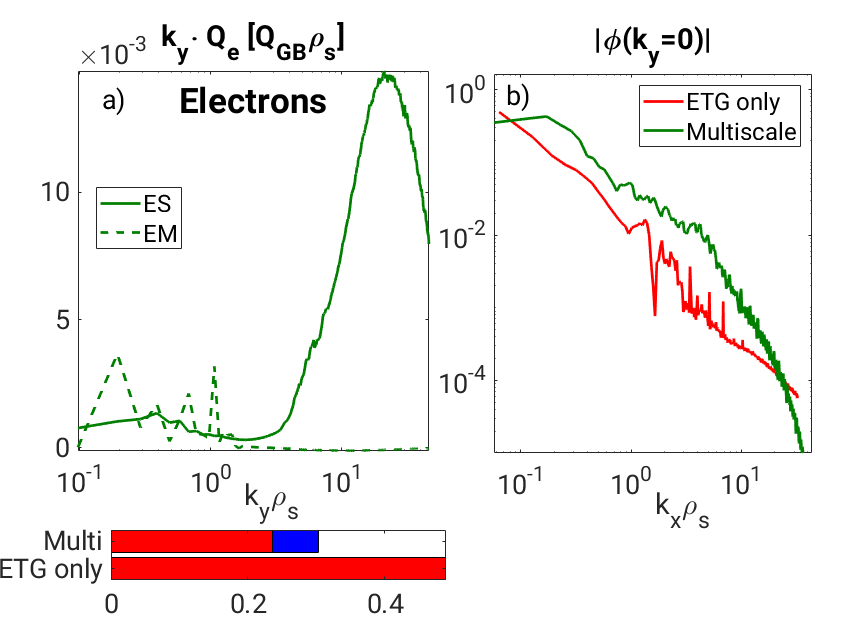}
    \caption{$(a)$ Electron heat fluxes multiplied by $k_y$ from multiscale simulations with $a/L_{T_i}=0$. Solid lines show electrostatic, dashed lines electromagnetic components. $(b)$ Radial spectra of the zonal flow $\phi(k_y=0)$ from multiscale and ETG-only simulations ($k_y>3$). Including the full ion-scale range captures marginally unstable TEMs/MTMs, which contribute little to heat flux but significantly to zonal field spectra.}
    \label{fig:5}
\end{figure}

\textit{Weak ITG Conditions.} -- In Fig.~\ref{fig:5}, we examine the opposite limit of vanishing ITG drive by setting $a/L_{T_i}=0$. Such a scenario can occur under dominant electron heating and high $\beta$, or strong $\bf E\times B$ shear flows, e.g. in the presence of a root transition \cite{Helander_Goodman_Beidler_Kuczynski_Smith_2024} suppressing the ITG.  While the linear spectra for this case (blue circles in Fig.~\ref{fig:1}) are nearly identical to those of the reference scenario, the nonlinear state differs markedly. As expected, transport is dominated by electron-scale dynamics, and the flux is almost entirely carried by ETG modes. However, the total electron heat flux is nearly halved in the multiscale simulation compared to the ETG-only case (note that from Fig.~\ref{fig:5} the flux spectra are multiplied by $k_y$ such that the area below the curve is proportional the total flux, allowing a visual comparison between  contributions from different scale). This reduction arises from cross-scale coupling via ion-scale zonal flows, as seen in the zonal spectra shown in Fig.~\ref{fig:5}b. Despite carrying only $\sim10\%$ of the total heat flux, the ion-scale modes -- primarily trapped-electron modes (TEMs) and MTMs -- significantly modify the zonal field structure, which in turn regulates the saturation level of ETG turbulence. This scenario provides a clear example where a multiscale approach is essential, as the sum of scale-separated simulations fails to capture the correct transport. 


\begin{figure}[h!]
    \centering
    \includegraphics[width=\linewidth]{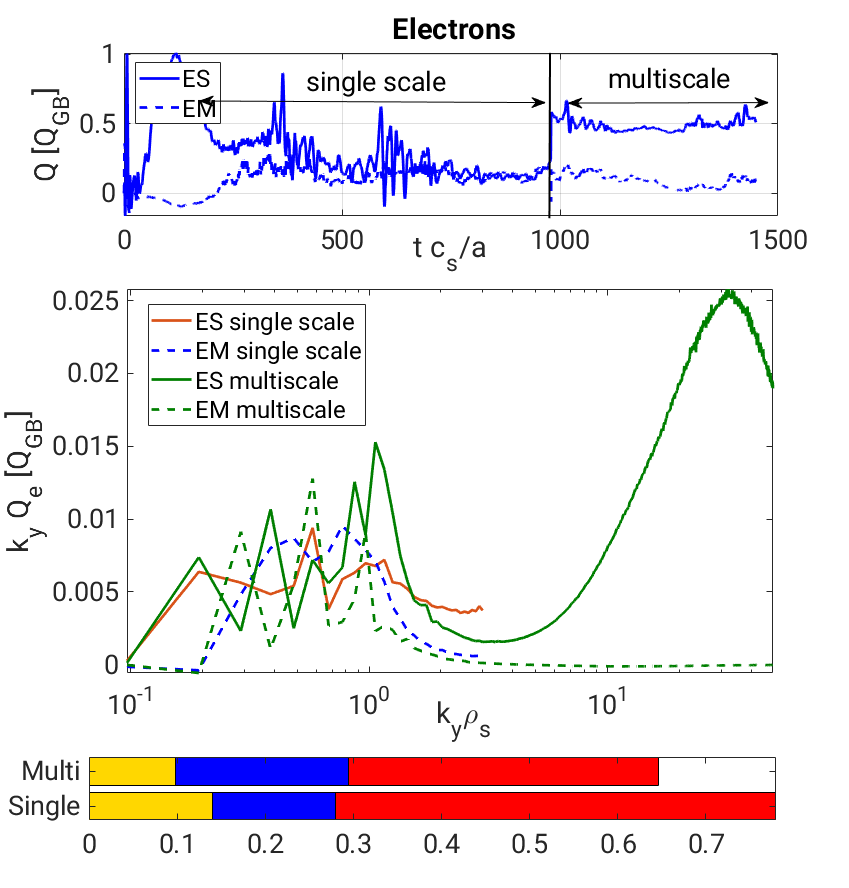}
    \caption{$(a)$ Time evolution of the electrostatic (solid lines) and electromagnetic (dashed) component if the electron  heat flux considering parameters where MTMs at the ion scale coexist with ETGs. A low-$k_y$ simulation (resolving only $k_y\rho_s <3$) is run up to 1000 $a/c_s$, and then continued with finer grids ($k_y\rho_s<50$) in order to include also ETGs. $(b)$  Heat flux spectra from scale-separated and multiscale simulations. Integrated contributions are indicated: electromagnetic (yellow), ion-scale electrostatic ($k_y\rho_s<3$, blue) and electron-scale (red).}
\label{fig:6}
\end{figure}

{\it Electromagnetic Scenarios.} -- The MTMs observed in the zero-ITG case above are not artifacts of our parameter choices but are expected features of high-$\beta$ reactor configurations. In fact, MTMs may also emerge in the W7-X core when collisionality is low enough to drive -- rather than dampen -- the instability, and when ITGs are suppressed by sufficiently high  beta. As a final multiscale scenario relevant to stellarators, we therefore consider a parameter set representative of such high-$\beta$ discharges. We set $a/L_{T_i}=1$, $a/L_n=2$, and  $a/L_{T_e}=6$. In this regime, MTMs are active at ion scales while ETGs remain unstable at high-$k$. Multiscale interactions between these two modes have been previously explored in tokamak core plasmas~\cite{Maeyama_PRL_119}, where ETG streamers were shown to disrupt the current layers that drive MTMs, effectively suppressing them.  However, the isotropic nature of stellarator ETGs suggests that cross-scale coupling may differ. Dedicated multiscale simulations are thus required. These runs are not only relevant to stellarators, but also to tokamak edge plasmas, where both MTMs and isotropic ETG turbulence have been reported.

For this case, we employ the same numerical setup as before, with increased binormal resolution to include 512 fully de-aliased $k_y$ modes (up to $k_y\rho_s=50$). Results for the electron heat flux (the ion channel is negligible due to ITG suppression) are summarized in Fig.~\ref{fig:6}. 

When restricting the analysis to ion-scale modes, the electron heat transport is found to be equally shared between electrostatic and electromagnetic components, with the latter peaking near $k_y\rho_s\sim1$, consistent with a mix of TEMs and MTMs. Unlike in tokamaks, MTMs in this stellarator scenario persist despite the presence of electron-scale modes. The electromagnetic heat flux decreases when all scales are include, dropping from 1.4 $Q_{\rm GB}$ to 1.0 $Q_{\rm GB}$, but is not fully suppressed. We interpret this as a consequence of the absence of strong ETG streamers, which in tokamaks can disrupt the current layers sustaining the MTMs. This interpretation aligns with results from~\cite{Pueschel_NF_2020}, where MTM–ETG interactions were studied in a simplified tokamak pedestal model.

Nevertheless, cross-scale effects remain critical in determining the total transport level: as shown in Fig.~\ref{fig:6}b, ETG-driven heat flux is overestimated by $\sim20\%$ when using a scale-separated approach. This underlines the importance of multiscale treatment in scenarios where both electromagnetic ion-scale and electron-scale modes coexist.

\textit{Summary and Discussion.} -- In this Letter, we have presented the first gyrokinetic simulations of multiscale turbulence in a stellarator, using the magnetic geometry of W7-X and experimentally relevant parameters. We explored a broad range of scenarios, including regimes where ETG turbulence coexists with varying levels of ITG turbulence, as well as cases involving MTMs relevant to high-$\beta$ and reactor-like conditions.

Notably, while ETG turbulence does not form the pronounced radial streamers observed in tokamaks, it can still drive substantial transport and interact with ion-scale turbulence. In regimes dominated by electrostatic ITG or TEM modes, cross-scale coupling closely resembles that in tokamaks: electron-scale fluctuations degrade zonal flows, thereby enhancing ion-scale transport, while ion-scale turbulence suppresses ETG activity. However, in scenarios where electromagnetic MTMs dominate at ion scales, the isotropic nature of ETG turbulence limits its ability to suppress them. As a result, MTMs remain robust even in the presence of multiscale interactions.

In all examined cases, both the magnitude and the spectral distribution of the turbulent heat flux differ substantially between multiscale simulations and the sum of scale-separated runs. This highlights the importance of resolving cross-scale dynamics to accurately predict transport levels in stellarator plasmas. Extending these results will require exploring a wider set of plasma profiles to develop simplified criteria -- similar in spirit to Ref.~\cite{Creely_2019} -- that can predict when multiscale effects are important. Establishing such guidelines based on single-scale linear or nonlinear simulations is left for future work.

{\it Acknowledgements.} - We acknowledge the EuroHPC Joint Undertaking for awarding this project access to the EuroHPC supercomputer LUMI, hosted by CSC (Finland) and the LUMI consortium through a EuroHPC Regular Access call.


\begin{thebibliography}{32}%
\makeatletter
\providecommand \@ifxundefined [1]{%
 \@ifx{#1\undefined}
}%
\providecommand \@ifnum [1]{%
 \ifnum #1\expandafter \@firstoftwo
 \else \expandafter \@secondoftwo
 \fi
}%
\providecommand \@ifx [1]{%
 \ifx #1\expandafter \@firstoftwo
 \else \expandafter \@secondoftwo
 \fi
}%
\providecommand \natexlab [1]{#1}%
\providecommand \enquote  [1]{``#1''}%
\providecommand \bibnamefont  [1]{#1}%
\providecommand \bibfnamefont [1]{#1}%
\providecommand \citenamefont [1]{#1}%
\providecommand \href@noop [0]{\@secondoftwo}%
\providecommand \href [0]{\begingroup \@sanitize@url \@href}%
\providecommand \@href[1]{\@@startlink{#1}\@@href}%
\providecommand \@@href[1]{\endgroup#1\@@endlink}%
\providecommand \@sanitize@url [0]{\catcode `\\12\catcode `\$12\catcode
  `\&12\catcode `\#12\catcode `\^12\catcode `\_12\catcode `\%12\relax}%
\providecommand \@@startlink[1]{}%
\providecommand \@@endlink[0]{}%
\providecommand \url  [0]{\begingroup\@sanitize@url \@url }%
\providecommand \@url [1]{\endgroup\@href {#1}{\urlprefix }}%
\providecommand \urlprefix  [0]{URL }%
\providecommand \Eprint [0]{\href }%
\providecommand \doibase [0]{https://doi.org/}%
\providecommand \selectlanguage [0]{\@gobble}%
\providecommand \bibinfo  [0]{\@secondoftwo}%
\providecommand \bibfield  [0]{\@secondoftwo}%
\providecommand \translation [1]{[#1]}%
\providecommand \BibitemOpen [0]{}%
\providecommand \bibitemStop [0]{}%
\providecommand \bibitemNoStop [0]{.\EOS\space}%
\providecommand \EOS [0]{\spacefactor3000\relax}%
\providecommand \BibitemShut  [1]{\csname bibitem#1\endcsname}%
\let\auto@bib@innerbib\@empty
\bibitem [{\citenamefont {Jenko}\ \emph {et~al.}(2000)\citenamefont {Jenko},
  \citenamefont {Dorland}, \citenamefont {Kotschenreuther},\ and\ \citenamefont
  {Rogers}}]{Jenko_POP_2010}%
  \BibitemOpen
  \bibfield  {author} {\bibinfo {author} {\bibfnamefont {F.}~\bibnamefont
  {Jenko}}, \bibinfo {author} {\bibfnamefont {W.}~\bibnamefont {Dorland}},
  \bibinfo {author} {\bibfnamefont {M.}~\bibnamefont {Kotschenreuther}},\ and\
  \bibinfo {author} {\bibfnamefont {B.~N.}\ \bibnamefont {Rogers}},\ }\bibfield
   {title} {\bibinfo {title} {Electron temperature gradient driven
  turbulence},\ }\href {https://doi.org/10.1063/1.874014} {\bibfield  {journal}
  {\bibinfo  {journal} {Physics of Plasmas}\ }\textbf {\bibinfo {volume} {7}},\
  \bibinfo {pages} {1904} (\bibinfo {year} {2000})}\BibitemShut {NoStop}%
\bibitem [{\citenamefont {Dorland}\ \emph {et~al.}(2000)\citenamefont
  {Dorland}, \citenamefont {Jenko}, \citenamefont {Kotschenreuther},\ and\
  \citenamefont {Rogers}}]{dorland_prl}%
  \BibitemOpen
  \bibfield  {author} {\bibinfo {author} {\bibfnamefont {W.}~\bibnamefont
  {Dorland}}, \bibinfo {author} {\bibfnamefont {F.}~\bibnamefont {Jenko}},
  \bibinfo {author} {\bibfnamefont {M.}~\bibnamefont {Kotschenreuther}},\ and\
  \bibinfo {author} {\bibfnamefont {B.~N.}\ \bibnamefont {Rogers}},\ }\bibfield
   {title} {\bibinfo {title} {Electron temperature gradient turbulence},\
  }\href {https://doi.org/10.1103/PhysRevLett.85.5579} {\bibfield  {journal}
  {\bibinfo  {journal} {Phys. Rev. Lett.}\ }\textbf {\bibinfo {volume} {85}},\
  \bibinfo {pages} {5579} (\bibinfo {year} {2000})}\BibitemShut {NoStop}%
\bibitem [{\citenamefont {Jenko}(2004)}]{jenko_2004}%
  \BibitemOpen
  \bibfield  {author} {\bibinfo {author} {\bibfnamefont {F.}~\bibnamefont
  {Jenko}},\ }\bibfield  {title} {\bibinfo {title} {On the nature of etg
  turbulence and cross-scale coupling},\ }\href@noop {} {\bibfield  {journal}
  {\bibinfo  {journal} {Journal of Plasma and Fusion Research}\ }\textbf
  {\bibinfo {volume} {6}} (\bibinfo {year} {2004})}\BibitemShut {NoStop}%
\bibitem [{\citenamefont {{Ren}}\ \emph {et~al.}(2024)\citenamefont {{Ren}},
  \citenamefont {{Guttenfelder}}, \citenamefont {{Kaye}},\ and\ \citenamefont
  {{Wang}}}]{RenRevModPhys2024}%
  \BibitemOpen
  \bibfield  {author} {\bibinfo {author} {\bibfnamefont {Y.}~\bibnamefont
  {{Ren}}}, \bibinfo {author} {\bibfnamefont {W.}~\bibnamefont
  {{Guttenfelder}}}, \bibinfo {author} {\bibfnamefont {S.~M.}\ \bibnamefont
  {{Kaye}}},\ and\ \bibinfo {author} {\bibfnamefont {W.~X.}\ \bibnamefont
  {{Wang}}},\ }\bibfield  {title} {\bibinfo {title} {{Transport from
  electron-scale turbulence in toroidal magnetic confinement devices}},\ }\href
  {https://doi.org/10.1007/s41614-023-00138-z} {\ \textbf {\bibinfo {volume}
  {8}},\ \bibinfo {eid} {5} (\bibinfo {year} {2024})}\BibitemShut {NoStop}%
\bibitem [{\citenamefont {Maeyama}\ \emph {et~al.}(2024)\citenamefont
  {Maeyama}, \citenamefont {Howard}, \citenamefont {Citrin}, \citenamefont
  {Watanabe},\ and\ \citenamefont {Tokuzawa}}]{Maeyama_NF_2024}%
  \BibitemOpen
  \bibfield  {author} {\bibinfo {author} {\bibfnamefont {S.}~\bibnamefont
  {Maeyama}}, \bibinfo {author} {\bibfnamefont {N.}~\bibnamefont {Howard}},
  \bibinfo {author} {\bibfnamefont {J.}~\bibnamefont {Citrin}}, \bibinfo
  {author} {\bibfnamefont {T.-H.}\ \bibnamefont {Watanabe}},\ and\ \bibinfo
  {author} {\bibfnamefont {T.}~\bibnamefont {Tokuzawa}},\ }\bibfield  {title}
  {\bibinfo {title} {Overview of multiscale turbulence studies covering
  ion-to-electron scales in magnetically confined fusion plasma},\ }\href
  {https://doi.org/10.1088/1741-4326/ad34e1} {\bibfield  {journal} {\bibinfo
  {journal} {Nuclear Fusion}\ }\textbf {\bibinfo {volume} {64}},\ \bibinfo
  {pages} {112007} (\bibinfo {year} {2024})}\BibitemShut {NoStop}%
\bibitem [{\citenamefont {Kotschenreuther}\ \emph {et~al.}(2019)\citenamefont
  {Kotschenreuther}, \citenamefont {Liu}, \citenamefont {Hatch}, \citenamefont
  {Mahajan}, \citenamefont {Zheng}, \citenamefont {Diallo}, \citenamefont
  {Groebner}, \citenamefont {the TEAM}, \citenamefont {Hillesheim},
  \citenamefont {Maggi}, \citenamefont {Giroud}, \citenamefont {Koechl},
  \citenamefont {Parail}, \citenamefont {Saarelma}, \citenamefont {Solano},
  \citenamefont {Chankin},\ and\ \citenamefont
  {Contributors}}]{Kotschenreuther_2019}%
  \BibitemOpen
  \bibfield  {author} {\bibinfo {author} {\bibfnamefont {M.}~\bibnamefont
  {Kotschenreuther}}, \bibinfo {author} {\bibfnamefont {X.}~\bibnamefont
  {Liu}}, \bibinfo {author} {\bibfnamefont {D.}~\bibnamefont {Hatch}}, \bibinfo
  {author} {\bibfnamefont {S.}~\bibnamefont {Mahajan}}, \bibinfo {author}
  {\bibfnamefont {L.}~\bibnamefont {Zheng}}, \bibinfo {author} {\bibfnamefont
  {A.}~\bibnamefont {Diallo}}, \bibinfo {author} {\bibfnamefont
  {R.}~\bibnamefont {Groebner}}, \bibinfo {author} {\bibfnamefont {D.-D.}\
  \bibnamefont {the TEAM}}, \bibinfo {author} {\bibfnamefont {J.}~\bibnamefont
  {Hillesheim}}, \bibinfo {author} {\bibfnamefont {C.}~\bibnamefont {Maggi}},
  \bibinfo {author} {\bibfnamefont {C.}~\bibnamefont {Giroud}}, \bibinfo
  {author} {\bibfnamefont {F.}~\bibnamefont {Koechl}}, \bibinfo {author}
  {\bibfnamefont {V.}~\bibnamefont {Parail}}, \bibinfo {author} {\bibfnamefont
  {S.}~\bibnamefont {Saarelma}}, \bibinfo {author} {\bibfnamefont
  {E.}~\bibnamefont {Solano}}, \bibinfo {author} {\bibfnamefont
  {A.}~\bibnamefont {Chankin}},\ and\ \bibinfo {author} {\bibfnamefont
  {J.}~\bibnamefont {Contributors}},\ }\bibfield  {title} {\bibinfo {title}
  {Gyrokinetic analysis and simulation of pedestals to identify the culprits
  for energy losses using ‘fingerprints’},\ }\href
  {https://doi.org/10.1088/1741-4326/ab1fa2} {\bibfield  {journal} {\bibinfo
  {journal} {Nuclear Fusion}\ }\textbf {\bibinfo {volume} {59}},\ \bibinfo
  {pages} {096001} (\bibinfo {year} {2019})}\BibitemShut {NoStop}%
\bibitem [{\citenamefont {Howard}\ \emph {et~al.}(2015)\citenamefont {Howard},
  \citenamefont {Holland}, \citenamefont {White}, \citenamefont {Greenwald},\
  and\ \citenamefont {Candy}}]{Howard_2016}%
  \BibitemOpen
  \bibfield  {author} {\bibinfo {author} {\bibfnamefont {N.}~\bibnamefont
  {Howard}}, \bibinfo {author} {\bibfnamefont {C.}~\bibnamefont {Holland}},
  \bibinfo {author} {\bibfnamefont {A.}~\bibnamefont {White}}, \bibinfo
  {author} {\bibfnamefont {M.}~\bibnamefont {Greenwald}},\ and\ \bibinfo
  {author} {\bibfnamefont {J.}~\bibnamefont {Candy}},\ }\bibfield  {title}
  {\bibinfo {title} {Multi-scale gyrokinetic simulation of tokamak plasmas:
  enhanced heat loss due to cross-scale coupling of plasma turbulence},\ }\href
  {https://doi.org/10.1088/0029-5515/56/1/014004} {\bibfield  {journal}
  {\bibinfo  {journal} {Nuclear Fusion}\ }\textbf {\bibinfo {volume} {56}},\
  \bibinfo {pages} {014004} (\bibinfo {year} {2015})}\BibitemShut {NoStop}%
\bibitem [{\citenamefont {Belli}\ \emph {et~al.}(2022)\citenamefont {Belli},
  \citenamefont {Candy},\ and\ \citenamefont {Sfiligoi}}]{Belli_2023}%
  \BibitemOpen
  \bibfield  {author} {\bibinfo {author} {\bibfnamefont {E.~A.}\ \bibnamefont
  {Belli}}, \bibinfo {author} {\bibfnamefont {J.}~\bibnamefont {Candy}},\ and\
  \bibinfo {author} {\bibfnamefont {I.}~\bibnamefont {Sfiligoi}},\ }\bibfield
  {title} {\bibinfo {title} {Spectral transition of multiscale turbulence in
  the tokamak pedestal},\ }\href {https://doi.org/10.1088/1361-6587/aca9fa}
  {\bibfield  {journal} {\bibinfo  {journal} {Plasma Physics and Controlled
  Fusion}\ }\textbf {\bibinfo {volume} {65}},\ \bibinfo {pages} {024001}
  (\bibinfo {year} {2022})}\BibitemShut {NoStop}%
\bibitem [{\citenamefont {Maeyama}\ \emph {et~al.}(2022)\citenamefont
  {Maeyama}, \citenamefont {Watanabe}, \citenamefont {Nakata}, \citenamefont
  {Nunami}, \citenamefont {Asahi},\ and\ \citenamefont
  {Ishizawa}}]{Maeyama2022}%
  \BibitemOpen
  \bibfield  {author} {\bibinfo {author} {\bibfnamefont {S.}~\bibnamefont
  {Maeyama}}, \bibinfo {author} {\bibfnamefont {T.-H.}\ \bibnamefont
  {Watanabe}}, \bibinfo {author} {\bibfnamefont {M.}~\bibnamefont {Nakata}},
  \bibinfo {author} {\bibfnamefont {M.}~\bibnamefont {Nunami}}, \bibinfo
  {author} {\bibfnamefont {Y.}~\bibnamefont {Asahi}},\ and\ \bibinfo {author}
  {\bibfnamefont {A.}~\bibnamefont {Ishizawa}},\ }\bibfield  {title} {\bibinfo
  {title} {Multi-scale turbulence simulation suggesting improvement of electron
  heated plasma confinement},\ }\href
  {https://doi.org/10.1038/s41467-022-30852-0} {\bibfield  {journal} {\bibinfo
  {journal} {Nature Communications}\ }\textbf {\bibinfo {volume} {13}},\
  \bibinfo {pages} {3166} (\bibinfo {year} {2022})}\BibitemShut {NoStop}%
\bibitem [{\citenamefont {Hatch}\ \emph {et~al.}(2024)\citenamefont {Hatch},
  \citenamefont {Kotschenreuther}, \citenamefont {Li}, \citenamefont
  {Chapman-Oplopoiou}, \citenamefont {Parisi}, \citenamefont {Mahajan},\ and\
  \citenamefont {Groebner}}]{Hatch_2024}%
  \BibitemOpen
  \bibfield  {author} {\bibinfo {author} {\bibfnamefont {D.}~\bibnamefont
  {Hatch}}, \bibinfo {author} {\bibfnamefont {M.}~\bibnamefont
  {Kotschenreuther}}, \bibinfo {author} {\bibfnamefont {P.-Y.}\ \bibnamefont
  {Li}}, \bibinfo {author} {\bibfnamefont {B.}~\bibnamefont
  {Chapman-Oplopoiou}}, \bibinfo {author} {\bibfnamefont {J.}~\bibnamefont
  {Parisi}}, \bibinfo {author} {\bibfnamefont {S.}~\bibnamefont {Mahajan}},\
  and\ \bibinfo {author} {\bibfnamefont {R.}~\bibnamefont {Groebner}},\
  }\bibfield  {title} {\bibinfo {title} {Modeling electron temperature profiles
  in the pedestal with simple formulas for etg transport},\ }\href
  {https://doi.org/10.1088/1741-4326/ad3972} {\bibfield  {journal} {\bibinfo
  {journal} {Nuclear Fusion}\ }\textbf {\bibinfo {volume} {64}},\ \bibinfo
  {pages} {066007} (\bibinfo {year} {2024})}\BibitemShut {NoStop}%
\bibitem [{\citenamefont {Jenko}\ and\ \citenamefont
  {Kendl}(2002)}]{Jenko_kendl_2002}%
  \BibitemOpen
  \bibfield  {author} {\bibinfo {author} {\bibfnamefont {F.}~\bibnamefont
  {Jenko}}\ and\ \bibinfo {author} {\bibfnamefont {A.}~\bibnamefont {Kendl}},\
  }\bibfield  {title} {\bibinfo {title} {Stellarator turbulence at electron
  gyroradius scales},\ }\href {https://doi.org/10.1088/1367-2630/4/1/335}
  {\bibfield  {journal} {\bibinfo  {journal} {New Journal of Physics}\ }\textbf
  {\bibinfo {volume} {4}},\ \bibinfo {pages} {35} (\bibinfo {year}
  {2002})}\BibitemShut {NoStop}%
\bibitem [{\citenamefont {Plunk}\ \emph {et~al.}(2019)\citenamefont {Plunk},
  \citenamefont {Xanthopoulos}, \citenamefont {Weir}, \citenamefont
  {Bozhenkov}, \citenamefont {Dinklage}, \citenamefont {Fuchert}, \citenamefont
  {Geiger}, \citenamefont {Hirsch}, \citenamefont {Hoefel}, \citenamefont
  {Jakubowski}, \citenamefont {Langenberg}, \citenamefont {Pablant},
  \citenamefont {Pasch}, \citenamefont {Stange}, \citenamefont {Zhang},\ and\
  \citenamefont {W7-X~Team}}]{Plunk_PRL_2019}%
  \BibitemOpen
  \bibfield  {author} {\bibinfo {author} {\bibfnamefont {G.~G.}\ \bibnamefont
  {Plunk}}, \bibinfo {author} {\bibfnamefont {P.}~\bibnamefont {Xanthopoulos}},
  \bibinfo {author} {\bibfnamefont {G.~M.}\ \bibnamefont {Weir}}, \bibinfo
  {author} {\bibfnamefont {S.~A.}\ \bibnamefont {Bozhenkov}}, \bibinfo {author}
  {\bibfnamefont {A.}~\bibnamefont {Dinklage}}, \bibinfo {author}
  {\bibfnamefont {G.}~\bibnamefont {Fuchert}}, \bibinfo {author} {\bibfnamefont
  {J.}~\bibnamefont {Geiger}}, \bibinfo {author} {\bibfnamefont
  {M.}~\bibnamefont {Hirsch}}, \bibinfo {author} {\bibfnamefont
  {U.}~\bibnamefont {Hoefel}}, \bibinfo {author} {\bibfnamefont
  {M.}~\bibnamefont {Jakubowski}}, \bibinfo {author} {\bibfnamefont
  {A.}~\bibnamefont {Langenberg}}, \bibinfo {author} {\bibfnamefont
  {N.}~\bibnamefont {Pablant}}, \bibinfo {author} {\bibfnamefont
  {E.}~\bibnamefont {Pasch}}, \bibinfo {author} {\bibfnamefont
  {T.}~\bibnamefont {Stange}}, \bibinfo {author} {\bibfnamefont
  {D.}~\bibnamefont {Zhang}},\ and\ \bibinfo {author} {\bibfnamefont
  {t.}~\bibnamefont {W7-X~Team}},\ }\bibfield  {title} {\bibinfo {title}
  {Stellarators resist turbulent transport on the electron larmor scale},\
  }\href {https://doi.org/10.1103/PhysRevLett.122.035002} {\bibfield  {journal}
  {\bibinfo  {journal} {Phys. Rev. Lett.}\ }\textbf {\bibinfo {volume} {122}},\
  \bibinfo {pages} {035002} (\bibinfo {year} {2019})}\BibitemShut {NoStop}%
\bibitem [{\citenamefont {Weir}\ \emph {et~al.}(2021)\citenamefont {Weir},
  \citenamefont {Xanthopoulos}, \citenamefont {Hirsch}, \citenamefont {Höfel},
  \citenamefont {Stange}, \citenamefont {Pablant}, \citenamefont {Grulke},
  \citenamefont {Äkäslompolo}, \citenamefont {Alcusón}, \citenamefont
  {Bozhenkov}, \citenamefont {Beurskens}, \citenamefont {Dinklage},
  \citenamefont {Fuchert}, \citenamefont {Geiger}, \citenamefont {Landreman},
  \citenamefont {Langenberg}, \citenamefont {Lazerson}, \citenamefont
  {Marushchenko}, \citenamefont {Pasch}, \citenamefont {Schilling},
  \citenamefont {Scott}, \citenamefont {Turkin}, \citenamefont {Klinger},\ and\
  \citenamefont {the Team}}]{Weir_2021}%
  \BibitemOpen
  \bibfield  {author} {\bibinfo {author} {\bibfnamefont {G.}~\bibnamefont
  {Weir}}, \bibinfo {author} {\bibfnamefont {P.}~\bibnamefont {Xanthopoulos}},
  \bibinfo {author} {\bibfnamefont {M.}~\bibnamefont {Hirsch}}, \bibinfo
  {author} {\bibfnamefont {U.}~\bibnamefont {Höfel}}, \bibinfo {author}
  {\bibfnamefont {T.}~\bibnamefont {Stange}}, \bibinfo {author} {\bibfnamefont
  {N.}~\bibnamefont {Pablant}}, \bibinfo {author} {\bibfnamefont
  {O.}~\bibnamefont {Grulke}}, \bibinfo {author} {\bibfnamefont
  {S.}~\bibnamefont {Äkäslompolo}}, \bibinfo {author} {\bibfnamefont
  {J.}~\bibnamefont {Alcusón}}, \bibinfo {author} {\bibfnamefont
  {S.}~\bibnamefont {Bozhenkov}}, \bibinfo {author} {\bibfnamefont
  {M.}~\bibnamefont {Beurskens}}, \bibinfo {author} {\bibfnamefont
  {A.}~\bibnamefont {Dinklage}}, \bibinfo {author} {\bibfnamefont
  {G.}~\bibnamefont {Fuchert}}, \bibinfo {author} {\bibfnamefont
  {J.}~\bibnamefont {Geiger}}, \bibinfo {author} {\bibfnamefont
  {M.}~\bibnamefont {Landreman}}, \bibinfo {author} {\bibfnamefont
  {A.}~\bibnamefont {Langenberg}}, \bibinfo {author} {\bibfnamefont
  {S.}~\bibnamefont {Lazerson}}, \bibinfo {author} {\bibfnamefont
  {N.}~\bibnamefont {Marushchenko}}, \bibinfo {author} {\bibfnamefont
  {E.}~\bibnamefont {Pasch}}, \bibinfo {author} {\bibfnamefont
  {J.}~\bibnamefont {Schilling}}, \bibinfo {author} {\bibfnamefont
  {E.}~\bibnamefont {Scott}}, \bibinfo {author} {\bibfnamefont
  {Y.}~\bibnamefont {Turkin}}, \bibinfo {author} {\bibfnamefont
  {T.}~\bibnamefont {Klinger}},\ and\ \bibinfo {author} {\bibfnamefont {W.-X.}\
  \bibnamefont {the Team}},\ }\bibfield  {title} {\bibinfo {title} {Heat pulse
  propagation and anomalous electron heat transport measurements on the
  optimized stellarator w7-x},\ }\href
  {https://doi.org/10.1088/1741-4326/abea55} {\bibfield  {journal} {\bibinfo
  {journal} {Nuclear Fusion}\ }\textbf {\bibinfo {volume} {61}},\ \bibinfo
  {pages} {056001} (\bibinfo {year} {2021})}\BibitemShut {NoStop}%
\bibitem [{\citenamefont {Wilms}\ \emph {et~al.}(2024)\citenamefont {Wilms},
  \citenamefont {{Bañón Navarro}}, \citenamefont {Windisch}, \citenamefont
  {Bozhenkov}, \citenamefont {Warmer}, \citenamefont {Fuchert}, \citenamefont
  {Ford}, \citenamefont {Zhang}, \citenamefont {Stange}, \citenamefont
  {Jenko},\ and\ \citenamefont {the Team}}]{Wilms_NF_2024}%
  \BibitemOpen
  \bibfield  {author} {\bibinfo {author} {\bibfnamefont {F.}~\bibnamefont
  {Wilms}}, \bibinfo {author} {\bibfnamefont {A.}~\bibnamefont {{Bañón
  Navarro}}}, \bibinfo {author} {\bibfnamefont {T.}~\bibnamefont {Windisch}},
  \bibinfo {author} {\bibfnamefont {S.}~\bibnamefont {Bozhenkov}}, \bibinfo
  {author} {\bibfnamefont {F.}~\bibnamefont {Warmer}}, \bibinfo {author}
  {\bibfnamefont {G.}~\bibnamefont {Fuchert}}, \bibinfo {author} {\bibfnamefont
  {O.}~\bibnamefont {Ford}}, \bibinfo {author} {\bibfnamefont {D.}~\bibnamefont
  {Zhang}}, \bibinfo {author} {\bibfnamefont {T.}~\bibnamefont {Stange}},
  \bibinfo {author} {\bibfnamefont {F.}~\bibnamefont {Jenko}},\ and\ \bibinfo
  {author} {\bibfnamefont {W.-X.}\ \bibnamefont {the Team}},\ }\bibfield
  {title} {\bibinfo {title} {Global gyrokinetic analysis of wendelstein 7-x
  discharge: unveiling the importance of trapped-electron-mode and
  electron-temperature-gradient turbulence},\ }\href
  {https://doi.org/10.1088/1741-4326/ad6675} {\bibfield  {journal} {\bibinfo
  {journal} {Nuclear Fusion}\ }\textbf {\bibinfo {volume} {64}},\ \bibinfo
  {pages} {096040} (\bibinfo {year} {2024})}\BibitemShut {NoStop}%
\bibitem [{\citenamefont {Agapito~Fernando}\ \emph {et~al.}(2025)\citenamefont
  {Agapito~Fernando}, \citenamefont {Bañón~Navarro}, \citenamefont
  {Carralero}, \citenamefont {Alonso}, \citenamefont {Di~Siena}, \citenamefont
  {Velasco}, \citenamefont {Wilms}, \citenamefont {Merlo}, \citenamefont
  {Jenko}, \citenamefont {Bozhenkov}, \citenamefont {Pasch}, \citenamefont
  {Fuchert}, \citenamefont {Brunner}, \citenamefont {Knauer}, \citenamefont
  {Langenberg}, \citenamefont {Pablant}, \citenamefont {Gonda}, \citenamefont
  {Ford}, \citenamefont {Vanó}, \citenamefont {Windisch}, \citenamefont
  {Estrada}, \citenamefont {Maragkoudakis},\ and\ \citenamefont {the
  Wendelstein 7-X~Team}}]{Fernando_PoP_2025}%
  \BibitemOpen
  \bibfield  {author} {\bibinfo {author} {\bibfnamefont {D.~L.~C.}\
  \bibnamefont {Agapito~Fernando}}, \bibinfo {author} {\bibfnamefont
  {A.}~\bibnamefont {Bañón~Navarro}}, \bibinfo {author} {\bibfnamefont
  {D.}~\bibnamefont {Carralero}}, \bibinfo {author} {\bibfnamefont
  {A.}~\bibnamefont {Alonso}}, \bibinfo {author} {\bibfnamefont
  {A.}~\bibnamefont {Di~Siena}}, \bibinfo {author} {\bibfnamefont {J.~L.}\
  \bibnamefont {Velasco}}, \bibinfo {author} {\bibfnamefont {F.}~\bibnamefont
  {Wilms}}, \bibinfo {author} {\bibfnamefont {G.}~\bibnamefont {Merlo}},
  \bibinfo {author} {\bibfnamefont {F.}~\bibnamefont {Jenko}}, \bibinfo
  {author} {\bibfnamefont {S.~A.}\ \bibnamefont {Bozhenkov}}, \bibinfo {author}
  {\bibfnamefont {E.}~\bibnamefont {Pasch}}, \bibinfo {author} {\bibfnamefont
  {G.}~\bibnamefont {Fuchert}}, \bibinfo {author} {\bibfnamefont {K.~J.}\
  \bibnamefont {Brunner}}, \bibinfo {author} {\bibfnamefont {J.}~\bibnamefont
  {Knauer}}, \bibinfo {author} {\bibfnamefont {A.}~\bibnamefont {Langenberg}},
  \bibinfo {author} {\bibfnamefont {N.~A.}\ \bibnamefont {Pablant}}, \bibinfo
  {author} {\bibfnamefont {T.}~\bibnamefont {Gonda}}, \bibinfo {author}
  {\bibfnamefont {O.~P.}\ \bibnamefont {Ford}}, \bibinfo {author}
  {\bibfnamefont {L.}~\bibnamefont {Vanó}}, \bibinfo {author} {\bibfnamefont
  {T.}~\bibnamefont {Windisch}}, \bibinfo {author} {\bibfnamefont
  {T.}~\bibnamefont {Estrada}}, \bibinfo {author} {\bibfnamefont
  {E.}~\bibnamefont {Maragkoudakis}},\ and\ \bibinfo {author} {\bibnamefont
  {the Wendelstein 7-X~Team}},\ }\bibfield  {title} {\bibinfo {title}
  {Validation of a comprehensive first-principles-based framework for
  predicting the performance of future stellarators},\ }\href
  {https://doi.org/10.1063/5.0267879} {\bibfield  {journal} {\bibinfo
  {journal} {Physics of Plasmas}\ }\textbf {\bibinfo {volume} {32}},\ \bibinfo
  {pages} {073904} (\bibinfo {year} {2025})}\BibitemShut {NoStop}%
\bibitem [{\citenamefont {{Germaschewski}}\ \emph {et~al.}(2021)\citenamefont
  {{Germaschewski}}, \citenamefont {{Allen}}, \citenamefont {{Dannert}},
  \citenamefont {{Hrywniak}}, \citenamefont {{Donaghy}}, \citenamefont
  {{Merlo}}, \citenamefont {{Ethier}}, \citenamefont {{D'Azevedo}},
  \citenamefont {{Jenko}},\ and\ \citenamefont
  {{Bhattacharjee}}}]{Germaschewski_PoP_2021}%
  \BibitemOpen
  \bibfield  {author} {\bibinfo {author} {\bibfnamefont {K.}~\bibnamefont
  {{Germaschewski}}}, \bibinfo {author} {\bibfnamefont {B.}~\bibnamefont
  {{Allen}}}, \bibinfo {author} {\bibfnamefont {T.}~\bibnamefont {{Dannert}}},
  \bibinfo {author} {\bibfnamefont {M.}~\bibnamefont {{Hrywniak}}}, \bibinfo
  {author} {\bibfnamefont {J.}~\bibnamefont {{Donaghy}}}, \bibinfo {author}
  {\bibfnamefont {G.}~\bibnamefont {{Merlo}}}, \bibinfo {author} {\bibfnamefont
  {S.}~\bibnamefont {{Ethier}}}, \bibinfo {author} {\bibfnamefont
  {E.}~\bibnamefont {{D'Azevedo}}}, \bibinfo {author} {\bibfnamefont
  {F.}~\bibnamefont {{Jenko}}},\ and\ \bibinfo {author} {\bibfnamefont
  {A.}~\bibnamefont {{Bhattacharjee}}},\ }\bibfield  {title} {\bibinfo {title}
  {{Toward exascale whole-device modeling of fusion devices: Porting the GENE
  gyrokinetic microturbulence code to GPU}},\ }\href
  {https://doi.org/10.1063/5.0046327} {\bibfield  {journal} {\bibinfo
  {journal} {Phys.~Plasmas}\ }\textbf {\bibinfo {volume} {28}},\ \bibinfo {eid}
  {062501} (\bibinfo {year} {2021})}\BibitemShut {NoStop}%
\bibitem [{\citenamefont {Maurer}\ \emph {et~al.}(2020)\citenamefont {Maurer},
  \citenamefont {{Bañón Navarro}}, \citenamefont {Dannert}, \citenamefont
  {Restelli}, \citenamefont {Hindenlang}, \citenamefont {Görler},
  \citenamefont {Told}, \citenamefont {Jarema}, \citenamefont {Merlo},\ and\
  \citenamefont {Jenko}}]{Maurer_JCP_2020}%
  \BibitemOpen
  \bibfield  {author} {\bibinfo {author} {\bibfnamefont {M.}~\bibnamefont
  {Maurer}}, \bibinfo {author} {\bibfnamefont {A.}~\bibnamefont {{Bañón
  Navarro}}}, \bibinfo {author} {\bibfnamefont {T.}~\bibnamefont {Dannert}},
  \bibinfo {author} {\bibfnamefont {M.}~\bibnamefont {Restelli}}, \bibinfo
  {author} {\bibfnamefont {F.}~\bibnamefont {Hindenlang}}, \bibinfo {author}
  {\bibfnamefont {T.}~\bibnamefont {Görler}}, \bibinfo {author} {\bibfnamefont
  {D.}~\bibnamefont {Told}}, \bibinfo {author} {\bibfnamefont {D.}~\bibnamefont
  {Jarema}}, \bibinfo {author} {\bibfnamefont {G.}~\bibnamefont {Merlo}},\ and\
  \bibinfo {author} {\bibfnamefont {F.}~\bibnamefont {Jenko}},\ }\bibfield
  {title} {\bibinfo {title} {Gene-3d: A global gyrokinetic turbulence code for
  stellarators},\ }\href {https://doi.org/10.1016/j.jcp.2020.109694} {\bibfield
   {journal} {\bibinfo  {journal} {Journal of Computational Physics}\ }\textbf
  {\bibinfo {volume} {420}},\ \bibinfo {pages} {109694} (\bibinfo {year}
  {2020})}\BibitemShut {NoStop}%
\bibitem [{\citenamefont {Martin}\ \emph {et~al.}(2018)\citenamefont {Martin},
  \citenamefont {Landreman}, \citenamefont {Xanthopoulos}, \citenamefont
  {Mandell},\ and\ \citenamefont {Dorland}}]{Martin_2018}%
  \BibitemOpen
  \bibfield  {author} {\bibinfo {author} {\bibfnamefont {M.~F.}\ \bibnamefont
  {Martin}}, \bibinfo {author} {\bibfnamefont {M.}~\bibnamefont {Landreman}},
  \bibinfo {author} {\bibfnamefont {P.}~\bibnamefont {Xanthopoulos}}, \bibinfo
  {author} {\bibfnamefont {N.~R.}\ \bibnamefont {Mandell}},\ and\ \bibinfo
  {author} {\bibfnamefont {W.}~\bibnamefont {Dorland}},\ }\bibfield  {title}
  {\bibinfo {title} {The parallel boundary condition for turbulence simulations
  in low magnetic shear devices},\ }\href
  {https://doi.org/10.1088/1361-6587/aad38a} {\bibfield  {journal} {\bibinfo
  {journal} {Plasma Physics and Controlled Fusion}\ }\textbf {\bibinfo {volume}
  {60}},\ \bibinfo {pages} {095008} (\bibinfo {year} {2018})}\BibitemShut
  {NoStop}%
\bibitem [{\citenamefont {Hindenlang}\ \emph {et~al.}(2025)\citenamefont
  {Hindenlang}, \citenamefont {Babin}, \citenamefont {Maj}, \citenamefont
  {Ribeiro}, \citenamefont {Koeberl}, \citenamefont {Muir}, \citenamefont
  {Plunk},\ and\ \citenamefont {Sonnendrücker}}]{gvec_2025_zenodo}%
  \BibitemOpen
  \bibfield  {author} {\bibinfo {author} {\bibfnamefont {F.}~\bibnamefont
  {Hindenlang}}, \bibinfo {author} {\bibfnamefont {R.}~\bibnamefont {Babin}},
  \bibinfo {author} {\bibfnamefont {O.}~\bibnamefont {Maj}}, \bibinfo {author}
  {\bibfnamefont {T.}~\bibnamefont {Ribeiro}}, \bibinfo {author} {\bibfnamefont
  {R.}~\bibnamefont {Koeberl}}, \bibinfo {author} {\bibfnamefont
  {D.}~\bibnamefont {Muir}}, \bibinfo {author} {\bibfnamefont {G.}~\bibnamefont
  {Plunk}},\ and\ \bibinfo {author} {\bibfnamefont {E.}~\bibnamefont
  {Sonnendrücker}},\ }\href {https://doi.org/10.5281/zenodo.15026781}
  {\bibinfo {title} {Gvec: A flexible 3d mhd equilibrium solver}} (\bibinfo
  {year} {2025})\BibitemShut {NoStop}%
\bibitem [{\citenamefont {Xanthopoulos}\ \emph {et~al.}(2020)\citenamefont
  {Xanthopoulos}, \citenamefont {Bozhenkov}, \citenamefont {Beurskens},
  \citenamefont {Smith}, \citenamefont {Plunk}, \citenamefont {Helander},
  \citenamefont {Beidler}, \citenamefont {Alcus\'on}, \citenamefont {Alonso},
  \citenamefont {Dinklage}, \citenamefont {Ford}, \citenamefont {Fuchert},
  \citenamefont {Geiger}, \citenamefont {Proll}, \citenamefont {Pueschel},
  \citenamefont {Turkin}, \citenamefont {Warmer},\ and\ \citenamefont
  {Team}}]{Xanthopoulos_PRL_2020}%
  \BibitemOpen
  \bibfield  {author} {\bibinfo {author} {\bibfnamefont {P.}~\bibnamefont
  {Xanthopoulos}}, \bibinfo {author} {\bibfnamefont {S.~A.}\ \bibnamefont
  {Bozhenkov}}, \bibinfo {author} {\bibfnamefont {M.~N.}\ \bibnamefont
  {Beurskens}}, \bibinfo {author} {\bibfnamefont {H.~M.}\ \bibnamefont
  {Smith}}, \bibinfo {author} {\bibfnamefont {G.~G.}\ \bibnamefont {Plunk}},
  \bibinfo {author} {\bibfnamefont {P.}~\bibnamefont {Helander}}, \bibinfo
  {author} {\bibfnamefont {C.~D.}\ \bibnamefont {Beidler}}, \bibinfo {author}
  {\bibfnamefont {J.~A.}\ \bibnamefont {Alcus\'on}}, \bibinfo {author}
  {\bibfnamefont {A.}~\bibnamefont {Alonso}}, \bibinfo {author} {\bibfnamefont
  {A.}~\bibnamefont {Dinklage}}, \bibinfo {author} {\bibfnamefont
  {O.}~\bibnamefont {Ford}}, \bibinfo {author} {\bibfnamefont {G.}~\bibnamefont
  {Fuchert}}, \bibinfo {author} {\bibfnamefont {J.}~\bibnamefont {Geiger}},
  \bibinfo {author} {\bibfnamefont {J.~H.~E.}\ \bibnamefont {Proll}}, \bibinfo
  {author} {\bibfnamefont {M.~J.}\ \bibnamefont {Pueschel}}, \bibinfo {author}
  {\bibfnamefont {Y.}~\bibnamefont {Turkin}}, \bibinfo {author} {\bibfnamefont
  {F.}~\bibnamefont {Warmer}},\ and\ \bibinfo {author} {\bibfnamefont
  {t.~W.-X.}\ \bibnamefont {Team}},\ }\bibfield  {title} {\bibinfo {title}
  {Turbulence mechanisms of enhanced performance stellarator plasmas},\ }\href
  {https://doi.org/10.1103/PhysRevLett.125.075001} {\bibfield  {journal}
  {\bibinfo  {journal} {Phys. Rev. Lett.}\ }\textbf {\bibinfo {volume} {125}},\
  \bibinfo {pages} {075001} (\bibinfo {year} {2020})}\BibitemShut {NoStop}%
\bibitem [{\citenamefont {Staebler}\ \emph {et~al.}(2016)\citenamefont
  {Staebler}, \citenamefont {Candy}, \citenamefont {Howard},\ and\
  \citenamefont {Holland}}]{Staebler_Pop_2016}%
  \BibitemOpen
  \bibfield  {author} {\bibinfo {author} {\bibfnamefont {G.~M.}\ \bibnamefont
  {Staebler}}, \bibinfo {author} {\bibfnamefont {J.}~\bibnamefont {Candy}},
  \bibinfo {author} {\bibfnamefont {N.~T.}\ \bibnamefont {Howard}},\ and\
  \bibinfo {author} {\bibfnamefont {C.}~\bibnamefont {Holland}},\ }\bibfield
  {title} {\bibinfo {title} {The role of zonal flows in the saturation of
  multi-scale gyrokinetic turbulence},\ }\href
  {https://doi.org/10.1063/1.4954905} {\bibfield  {journal} {\bibinfo
  {journal} {Physics of Plasmas}\ }\textbf {\bibinfo {volume} {23}},\ \bibinfo
  {pages} {062518} (\bibinfo {year} {2016})}\BibitemShut {NoStop}%
\bibitem [{\citenamefont {\textit{et al.}}(2016)}]{howard_multiscale_2016}%
  \BibitemOpen
  \bibfield  {author} {\bibinfo {author} {\bibfnamefont {N.~T.~H.}\
  \bibnamefont {\textit{et al.}}},\ }\bibfield  {title} {\bibinfo {title}
  {Multi-scale gyrokinetic simulations: {Comparison} with experiment and
  implications for predicting turbulence and transport},\ }\href
  {https://doi.org/10.1063/1.4946028} {\ \textbf {\bibinfo {volume} {23}},\
  \bibinfo {pages} {056109} (\bibinfo {year} {2016})}\BibitemShut {NoStop}%
\bibitem [{\citenamefont {Candy}\ \emph {et~al.}(2007)\citenamefont {Candy},
  \citenamefont {Waltz}, \citenamefont {Fahey},\ and\ \citenamefont
  {Holland}}]{Candy_2007}%
  \BibitemOpen
  \bibfield  {author} {\bibinfo {author} {\bibfnamefont {J.}~\bibnamefont
  {Candy}}, \bibinfo {author} {\bibfnamefont {R.~E.}\ \bibnamefont {Waltz}},
  \bibinfo {author} {\bibfnamefont {M.~R.}\ \bibnamefont {Fahey}},\ and\
  \bibinfo {author} {\bibfnamefont {C.}~\bibnamefont {Holland}},\ }\bibfield
  {title} {\bibinfo {title} {The effect of ion-scale dynamics on
  electron-temperature-gradient turbulence},\ }\href
  {https://doi.org/10.1088/0741-3335/49/8/008} {\bibfield  {journal} {\bibinfo
  {journal} {Plasma Physics and Controlled Fusion}\ }\textbf {\bibinfo {volume}
  {49}},\ \bibinfo {pages} {1209} (\bibinfo {year} {2007})}\BibitemShut
  {NoStop}%
\bibitem [{\citenamefont {G\"orler}\ and\ \citenamefont
  {Jenko}(2008)}]{Goerler_PRL_2008}%
  \BibitemOpen
  \bibfield  {author} {\bibinfo {author} {\bibfnamefont {T.}~\bibnamefont
  {G\"orler}}\ and\ \bibinfo {author} {\bibfnamefont {F.}~\bibnamefont
  {Jenko}},\ }\bibfield  {title} {\bibinfo {title} {Scale separation between
  electron and ion thermal transport},\ }\href
  {https://doi.org/10.1103/PhysRevLett.100.185002} {\bibfield  {journal}
  {\bibinfo  {journal} {Phys. Rev. Lett.}\ }\textbf {\bibinfo {volume} {100}},\
  \bibinfo {pages} {185002} (\bibinfo {year} {2008})}\BibitemShut {NoStop}%
\bibitem [{\citenamefont {Maeyama}\ \emph {et~al.}(2015)\citenamefont
  {Maeyama}, \citenamefont {Idomura}, \citenamefont {Watanabe}, \citenamefont
  {Nakata}, \citenamefont {Yagi}, \citenamefont {Miyato}, \citenamefont
  {Ishizawa},\ and\ \citenamefont {Nunami}}]{maeyama_cross-scale_2015}%
  \BibitemOpen
  \bibfield  {author} {\bibinfo {author} {\bibfnamefont {S.}~\bibnamefont
  {Maeyama}}, \bibinfo {author} {\bibfnamefont {Y.}~\bibnamefont {Idomura}},
  \bibinfo {author} {\bibfnamefont {T.-H.}\ \bibnamefont {Watanabe}}, \bibinfo
  {author} {\bibfnamefont {M.}~\bibnamefont {Nakata}}, \bibinfo {author}
  {\bibfnamefont {M.}~\bibnamefont {Yagi}}, \bibinfo {author} {\bibfnamefont
  {N.}~\bibnamefont {Miyato}}, \bibinfo {author} {\bibfnamefont
  {A.}~\bibnamefont {Ishizawa}},\ and\ \bibinfo {author} {\bibfnamefont
  {M.}~\bibnamefont {Nunami}},\ }\bibfield  {title} {\bibinfo {title}
  {Cross-scale interactions between electron and ion scale turbulence in a
  tokamak plasma},\ }\href {https://doi.org/10.1103/PhysRevLett.114.255002}
  {\bibfield  {journal} {\bibinfo  {journal} {Phys. Rev. Lett.}\ }\textbf
  {\bibinfo {volume} {114}},\ \bibinfo {pages} {255002} (\bibinfo {year}
  {2015})}\BibitemShut {NoStop}%
\bibitem [{\citenamefont {Howard}\ \emph {et~al.}(2016)\citenamefont {Howard},
  \citenamefont {Holland}, \citenamefont {White}, \citenamefont {Greenwald},
  \citenamefont {Candy},\ and\ \citenamefont {Creely}}]{Howard_PoP_2016}%
  \BibitemOpen
  \bibfield  {author} {\bibinfo {author} {\bibfnamefont {N.~T.}\ \bibnamefont
  {Howard}}, \bibinfo {author} {\bibfnamefont {C.}~\bibnamefont {Holland}},
  \bibinfo {author} {\bibfnamefont {A.~E.}\ \bibnamefont {White}}, \bibinfo
  {author} {\bibfnamefont {M.}~\bibnamefont {Greenwald}}, \bibinfo {author}
  {\bibfnamefont {J.}~\bibnamefont {Candy}},\ and\ \bibinfo {author}
  {\bibfnamefont {A.~J.}\ \bibnamefont {Creely}},\ }\bibfield  {title}
  {\bibinfo {title} {Multi-scale gyrokinetic simulations: Comparison with
  experiment and implications for predicting turbulence and transport},\ }\href
  {https://doi.org/10.1063/1.4946028} {\bibfield  {journal} {\bibinfo
  {journal} {Physics of Plasmas}\ }\textbf {\bibinfo {volume} {23}},\ \bibinfo
  {pages} {056109} (\bibinfo {year} {2016})}\BibitemShut {NoStop}%
\bibitem [{\citenamefont {Holland}\ \emph {et~al.}(2017)\citenamefont
  {Holland}, \citenamefont {Howard},\ and\ \citenamefont
  {Grierson}}]{Holland_2017}%
  \BibitemOpen
  \bibfield  {author} {\bibinfo {author} {\bibfnamefont {C.}~\bibnamefont
  {Holland}}, \bibinfo {author} {\bibfnamefont {N.}~\bibnamefont {Howard}},\
  and\ \bibinfo {author} {\bibfnamefont {B.}~\bibnamefont {Grierson}},\
  }\bibfield  {title} {\bibinfo {title} {Gyrokinetic predictions of multiscale
  transport in a diii-d iter baseline discharge},\ }\href
  {https://doi.org/10.1088/1741-4326/aa6c16} {\bibfield  {journal} {\bibinfo
  {journal} {Nuclear Fusion}\ }\textbf {\bibinfo {volume} {57}},\ \bibinfo
  {pages} {066043} (\bibinfo {year} {2017})}\BibitemShut {NoStop}%
\bibitem [{\citenamefont {Howard}\ \emph {et~al.}(2017)\citenamefont {Howard},
  \citenamefont {Holland}, \citenamefont {White}, \citenamefont {Greenwald},
  \citenamefont {Rodriguez-Fernandez}, \citenamefont {Candy},\ and\
  \citenamefont {Creely}}]{Howard_2018}%
  \BibitemOpen
  \bibfield  {author} {\bibinfo {author} {\bibfnamefont {N.~T.}\ \bibnamefont
  {Howard}}, \bibinfo {author} {\bibfnamefont {C.}~\bibnamefont {Holland}},
  \bibinfo {author} {\bibfnamefont {A.~E.}\ \bibnamefont {White}}, \bibinfo
  {author} {\bibfnamefont {M.}~\bibnamefont {Greenwald}}, \bibinfo {author}
  {\bibfnamefont {P.}~\bibnamefont {Rodriguez-Fernandez}}, \bibinfo {author}
  {\bibfnamefont {J.}~\bibnamefont {Candy}},\ and\ \bibinfo {author}
  {\bibfnamefont {A.~J.}\ \bibnamefont {Creely}},\ }\bibfield  {title}
  {\bibinfo {title} {Multi-scale gyrokinetic simulations of an alcator c-mod,
  elm-y h-mode plasma},\ }\href {https://doi.org/10.1088/1361-6587/aa9148}
  {\bibfield  {journal} {\bibinfo  {journal} {Plasma Physics and Controlled
  Fusion}\ }\textbf {\bibinfo {volume} {60}},\ \bibinfo {pages} {014034}
  (\bibinfo {year} {2017})}\BibitemShut {NoStop}%
\bibitem [{\citenamefont {Helander}\ \emph {et~al.}(2024)\citenamefont
  {Helander}, \citenamefont {Goodman}, \citenamefont {Beidler}, \citenamefont
  {Kuczynski},\ and\ \citenamefont
  {Smith}}]{Helander_Goodman_Beidler_Kuczynski_Smith_2024}%
  \BibitemOpen
  \bibfield  {author} {\bibinfo {author} {\bibfnamefont {P.}~\bibnamefont
  {Helander}}, \bibinfo {author} {\bibfnamefont {A.}~\bibnamefont {Goodman}},
  \bibinfo {author} {\bibfnamefont {C.}~\bibnamefont {Beidler}}, \bibinfo
  {author} {\bibfnamefont {M.}~\bibnamefont {Kuczynski}},\ and\ \bibinfo
  {author} {\bibfnamefont {H.}~\bibnamefont {Smith}},\ }\bibfield  {title}
  {\bibinfo {title} {Optimised stellarators with a positive radial electric
  field},\ }\href {https://doi.org/10.1017/S0022377824001004} {\bibfield
  {journal} {\bibinfo  {journal} {Journal of Plasma Physics}\ }\textbf
  {\bibinfo {volume} {90}},\ \bibinfo {pages} {175900602} (\bibinfo {year}
  {2024})}\BibitemShut {NoStop}%
\bibitem [{\citenamefont {Maeyama}\ \emph {et~al.}(2017)\citenamefont
  {Maeyama}, \citenamefont {Watanabe},\ and\ \citenamefont
  {Ishizawa}}]{Maeyama_PRL_119}%
  \BibitemOpen
  \bibfield  {author} {\bibinfo {author} {\bibfnamefont {S.}~\bibnamefont
  {Maeyama}}, \bibinfo {author} {\bibfnamefont {T.-H.}\ \bibnamefont
  {Watanabe}},\ and\ \bibinfo {author} {\bibfnamefont {A.}~\bibnamefont
  {Ishizawa}},\ }\bibfield  {title} {\bibinfo {title} {Suppression of ion-scale
  microtearing modes by electron-scale turbulence via cross-scale nonlinear
  interactions in tokamak plasmas},\ }\href
  {https://doi.org/10.1103/PhysRevLett.119.195002} {\bibfield  {journal}
  {\bibinfo  {journal} {Phys. Rev. Lett.}\ }\textbf {\bibinfo {volume} {119}},\
  \bibinfo {pages} {195002} (\bibinfo {year} {2017})}\BibitemShut {NoStop}%
\bibitem [{\citenamefont {Pueschel}\ \emph {et~al.}(2020)\citenamefont
  {Pueschel}, \citenamefont {Hatch}, \citenamefont {Kotschenreuther},
  \citenamefont {Ishizawa},\ and\ \citenamefont {Merlo}}]{Pueschel_NF_2020}%
  \BibitemOpen
  \bibfield  {author} {\bibinfo {author} {\bibfnamefont {M.}~\bibnamefont
  {Pueschel}}, \bibinfo {author} {\bibfnamefont {D.}~\bibnamefont {Hatch}},
  \bibinfo {author} {\bibfnamefont {M.}~\bibnamefont {Kotschenreuther}},
  \bibinfo {author} {\bibfnamefont {A.}~\bibnamefont {Ishizawa}},\ and\
  \bibinfo {author} {\bibfnamefont {G.}~\bibnamefont {Merlo}},\ }\bibfield
  {title} {\bibinfo {title} {Multi-scale interactions of microtearing
  turbulence in the tokamak pedestal},\ }\href
  {https://doi.org/10.1088/1741-4326/abba49} {\bibfield  {journal} {\bibinfo
  {journal} {Nuclear Fusion}\ }\textbf {\bibinfo {volume} {60}},\ \bibinfo
  {pages} {124005} (\bibinfo {year} {2020})}\BibitemShut {NoStop}%
\bibitem [{\citenamefont {Creely}\ \emph {et~al.}(2019)\citenamefont {Creely},
  \citenamefont {Rodriguez-Fernandez}, \citenamefont {Conway}, \citenamefont
  {Freethy}, \citenamefont {Howard}, \citenamefont {White},\ and\ \citenamefont
  {the Team}}]{Creely_2019}%
  \BibitemOpen
  \bibfield  {author} {\bibinfo {author} {\bibfnamefont {A.~J.}\ \bibnamefont
  {Creely}}, \bibinfo {author} {\bibfnamefont {P.}~\bibnamefont
  {Rodriguez-Fernandez}}, \bibinfo {author} {\bibfnamefont {G.~D.}\
  \bibnamefont {Conway}}, \bibinfo {author} {\bibfnamefont {S.~J.}\
  \bibnamefont {Freethy}}, \bibinfo {author} {\bibfnamefont {N.~T.}\
  \bibnamefont {Howard}}, \bibinfo {author} {\bibfnamefont {A.~E.}\
  \bibnamefont {White}},\ and\ \bibinfo {author} {\bibfnamefont {A.~U.}\
  \bibnamefont {the Team}},\ }\bibfield  {title} {\bibinfo {title} {Criteria
  for the importance of multi-scale interactions in turbulent transport
  simulations},\ }\href {https://doi.org/10.1088/1361-6587/ab24ae} {\bibfield
  {journal} {\bibinfo  {journal} {Plasma Physics and Controlled Fusion}\
  }\textbf {\bibinfo {volume} {61}},\ \bibinfo {pages} {085022} (\bibinfo
  {year} {2019})}\BibitemShut {NoStop}%
\end{thebibliography}
\end{document}